\documentclass[showpacs,preprintnumbers,amsmath,amssymb,twocolumn,nofootinbib]{revtex4}
\usepackage{graphicx}% Include figure files
\usepackage{dcolumn}% Align table columns on decimal point
\usepackage{bm}% bold math
\usepackage{hyperref}
\usepackage{epsfig}
\newcommand{\nco}{\newcommand}
\nco{\beq}{\begin{equation}} \nco{\eeq}{\end{equation}}
\nco{\beqa}{\begin{eqnarray}} \nco{\eeqa}{\end{eqnarray}}
\nco{\lra}{\leftrightarrow}
\def\sfrac#1#2{{\textstyle{#1\over #2}}}

\nco{\sss}{\scriptscriptstyle} \nco{\dphi}{\varphi}
\nco{\lsim}{\mbox{\raisebox{-.6ex}{~$\stackrel{<}{\sim}$~}}}
\nco{\gsim}{\mbox{\raisebox{-.6ex}{~$\stackrel{>}{\sim}$~}}}

\begin{document}

%\preprint{McGill 03-25}

\title{A new twist on excited dark matter: implications for
INTEGRAL, PAMELA/ATIC/PPB-BETS, DAMA}

\author{Fang Chen, James M.\ Cline, Andrew R.\ Frey}

\affiliation{%
\centerline{Physics Department, McGill University,
3600 University Street, Montr\'eal, Qu\'ebec, Canada H3A 2T8}
e-mail: fangchen, jcline, frey\ @physics.mcgill.ca }

\date{October 24, 2009 (version 5)}

\begin{abstract} 

We show that the 511 keV gamma ray excess observed by INTEGRAL/SPI
can be more robustly explained by exciting dark matter (DM) at the center
of the galaxy, if there is a peculiar spectrum of DM states $\chi_0$,
$\chi_1$ and $\chi_2$, with masses $M_0 \sim 500$ GeV, $M_1 \lsim M_0
+ 2 m_e$, and $M_2 = M_1 + \delta M \gsim M_0 + 2 m_e$.  The small
mass splitting $\delta M$ should be $\lsim 100$ keV. In addition, we 
require at least two new gauge bosons (preferably three), with
masses  $\sim\!\! 100$ MeV.   With this spectrum, $\chi_1$ is stable,
but can be excited to $\chi_2$ by low-velocity DM scatterings near
the galactic center, which are Sommerfeld-enhanced  by two of the 100
MeV gauge boson exchanges.  The excited state $\chi_2$ decays to
$\chi_0$ and nonrelativistic $e^+e^-$, mediated by the third gauge
boson, which mixes with the photon and $Z$.  Although such a small
100 keV splitting has been independently proposed for explaining the
DAMA annual modulation through the inelastic DM mechanism, the need
for stability of $\chi_1$ (and hence seqestering it from the Standard
Model) implies that our scenario {\it cannot} account for the DAMA
signal.   It can however address the PAMELA/ATIC positron excess via
DM annihilation in the galaxy, and it offers the possibility of a
sharper feature in the ATIC spectrum relative to previously proposed
models.  The data are consistent with three new gauge bosons,  whose
couplings fit naturally into a broken SU(2) gauge theory where the DM
is a triplet of the SU(2).  We propose a simple model in which the
SU(2) is broken by new Higgs triplet and 5-plet VEV's, giving rise
to the right spectrum of DM,  and mixing of one of the new gauge
bosons with the photon and $Z$ boson.  A coupling of the DM to a heavy
$Z'$ may also be necessary to get the right relic density and
PAMELA/ATIC signals.

\end{abstract}

\maketitle

\section{Introduction} Dramatic developments in observational astronomy have started to
alter our picture of dark matter (DM); instead of being a single
state, observations have indirectly suggested that DM could be a
multiplet with small mass splittings.  In ref.\ \cite{AH} it was
argued that such a scenario can explain not only the
positron/electron excess recently indicated in the
10-100 GeV region by the PAMELA
\cite{pamela} and (to some extent) HEAT \cite{heat} experiments,
and a similar one in
the 500-800 GeV region seen by  
ATIC \cite{atic} and PPB-BETS \cite{ppb-bets},
but also the 511 keV gamma rays observed by INTEGRAL/SPI
\cite{integral,integral2}, the WMAP haze \cite{haze}
and the annual modulation observed by DAMA/LIBRA
\cite{dama}.   The connections were further explored in \cite{cholis}.
The common link between them is that they can be explained in terms
of DM undergoing enhanced scattering and subsequent annihilation into
light bosons which decay to $e^+e^-$ \cite{secluded}, or else
exciting a DM state
with a small mass splitting above the ground state, which might decay
back to the ground state and $e^+e^-$ if the splitting is greater than
$2m_e$.  In the case of DAMA, the mass splitting
provides the kinematics  which would enable DAMA to be more sensitive
than other experiments to a necessarily inelastic collision.  The ideas of
excited dark matter (XDM) \cite{FW} and inelastic dark matter (IDM)
\cite{iDM} were proposed before ref.\ \cite{AH}, but the latter took
the step of trying to unify them into an appealing theoretical
framework, and to use it to also explain the excess positron/electron
results of PAMELA/ATIC.

The unified description of dark matter has one shortcoming, however.
In ref.\ \cite{PR}, it was shown that the XDM mechanism falls short
of being able to reproduce the experimental observation by nearly three
orders of magnitude, even if the galactic DM scattering cross section
$\sigma_{\rm gal}$ saturates the unitarity limit in the s-wave
contribution.  The main loophole for circumventing this conclusion
was to hope that higher-$l$ partial waves could increase $\sigma_{\rm
gal}$ by a factor of 300. Such an enhancement was argued to be unlikely in ref.\ 
\cite{PR}, and we will show that this argument is borne out in the
class of models proposed by ref.\ \cite{AH}.  Thus the XDM
explanation of the 511 keV excess remains unrealized.\footnote{ This
conclusion depends on the small error bars for the necessary value of
$\sigma_{\rm gal}$  estimated by ref.\ \cite{ldm2}.  However there
seems to be a larger uncertainty in the DM density at the center of
the galaxy, $n_c$. Since positron production in the XDM scenario
scales like $n_c^2$, the original XDM idea with $\delta M\cong m_e$
might be salvaged if $n_c$ is greater by a factor of 17 (even more if
the unitarity bound is not saturated) than in the model found to be
preferred in the best fit to the INTEGRAL data by ref.\
\cite{ldm2}.  It would be worthwhile, though beyond the scope of the
present work, to further investigate this point.\label{fnote1}}

We note that the 511 keV anomaly is not just the finding of the
INTEGRAL experiment, but it was first observed in 1972, and has been
seen in four subsequent balloon- and satellite-borne experiments
\cite{integral2}.  The observation is thus quite credible, and so far
is lacking any highly convincing astrophysical or particle physics
explanation, although attempts have been made using positron emission
from low-mass x-ray binaries (LMXB's) \cite{weid}, or  annihilation
of light MeV-scale dark matter (see for example \cite{ldm,ldm2}) as
well as decaying relics \cite{PR,decays}.  The LMXB hypothesis is
predicated on a supposed correlation between the asymmetry in the 
disk component of the 511 keV gamma rays and the distribution of
bright LMXB's, but this has been criticized on several grounds
in ref.\ \cite{silk} (although the discrepancies might be ameliorated
if positrons produced in the disk can be transported to the bulge
before annihilating \cite{lmxb}).  Therefore it is still 
interesting to find a technically natural particle physics
explanation for the 511 keV line, regardless of the other
experimental anomalies.  This was the primary motivation for the
present work.  The fact that our positive finding for the INTEGRAL
anomaly is consistent with the general framework outlined in ref.\
\cite{AH}, for also explaining the other experiments (apart from
DAMA), heightens its interest.

\begin{figure}[t]
\smallskip \centerline{\epsfxsize=0.45\textwidth\epsfbox{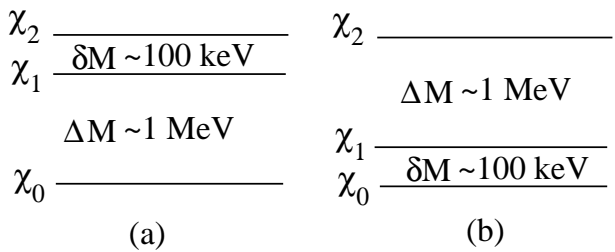}}
\caption{Left: DM spectrum needed in present work to account for the
INTEGRAL/SPI observations; right:
spectrum suggested by ref.\ \cite{AH}.}
\label{spect}
\end{figure}

\begin{figure}[t]
\smallskip
\centerline{\epsfxsize=0.45\textwidth\epsfbox{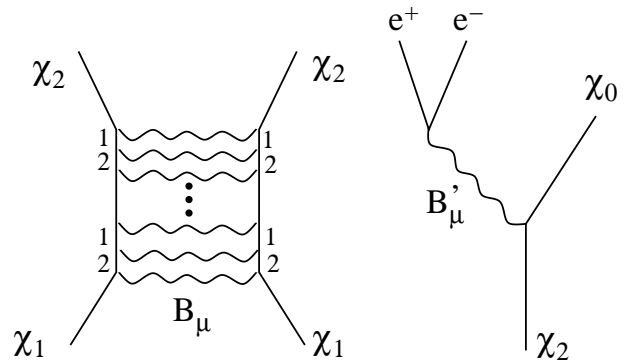}}
\caption{Left: Sommerfeld-enhanced scattering
$\chi_1\chi_1\to\chi_2\chi_2$.  Right: decay $\chi_2\to\chi_0 e^+
e_-$.}
\label{scatt}
\end{figure}

Our new twist for making XDM viable is to have a mass splitting
$\delta M$ which is much smaller than $m_e$ between the middle DM
state and the heaviest one, as 
illustrated in figure \ref{spect}. 
These are assumed to be Majorana fermions, $\chi_i$. We will design
the model so that $\chi_1$ cannot  decay directly to the ground state
$\chi_0$; thus both $\chi_0$ and $\chi_1$ are stable.  $\chi_1$
undergoes Sommerfeld-enhanced scattering \cite{AH,Hisano,LS,MR} at the galactic center
(where the DM density is highest) through multiple exchange of a
light gauge boson $B_\mu$ with off-diagonal coupling
$\bar\chi_2\gamma^\mu B_\mu \chi_1$, as envisioned in ref.\
\cite{AH}. The produced $\chi_2$ states subsequently decay through
another vector $B'_\mu$ which mixes with the photon to produce
$e^+e^-$.  These processes are shown in figure \ref{scatt}.  The key
ingredient which makes the scattering efficient enough is the small
mass splitting $\delta M$.  Previous attempts to implement the XDM
mechanism have failed because a larger splitting $\sim m_e$ was
assumed, and this makes the excitation rate too small.
The figure of
merit is the ratio of the predicted rate of $e^+$ production to that
observed, in the $l$th partial wave, assuming the
unitarity bound is saturated \cite{PR}:
\beq
	R_l = 4.5\times 10^{-4}\, {2l+1\over v_0}\left({500\ {\rm GeV}
	\over M}\right)^4 e^{-2m_e/Mv_0^2}
\label{rl}
\eeq
It is suppressed by the Maxwell-Boltzmann distribution at the
threshold for production of the excited state.  Eq.\ (\ref{rl})
was predicated on the assumption that
the kinetic energy of each DM particle must be sufficient to produce
one electron or positron.  Since the characteristic DM velocity
$v_0$ is fixed,  $M$ must be
sufficiently large to avoid the Boltzmann suppression, but the rate
also scales like $M^{-4}$.  Even with the optimal value of
$M=m_e/2v_0^2$, one
would need to more than double the estimated value of $v_0=6\times
10^{-4} c$
to make $R_0=1$.  Such a large change seems to be well outside the range of
uncertainty in
the current understanding of the DM velocity distribution.
 However in our scenario, $m_e$ is replaced by the
smaller $\delta M$, which can significantly ease this tension.  We
will show that $\sum_l R_l$ can be 1 as required without changing
$v_0$, and keeping $M\sim 500$ GeV as desired for PAMELA/ATIC, if
$\delta M \lsim 100$ keV.  It is intriguing that the same splitting 
has been advocated previously to account for the DAMA signal. However
we will show that the need for stability of $\chi_1$ means that it
cannot interact with baryons or leptons at detectable levels;
 thus our proposal does not seem
to be compatible with the IDM explanation for DAMA.

In the remainder of the paper we will give details of the computation
of the excitation rate (section II), tighten the case against the
large mass gap $\delta M=m_e$ scenario in section III, and show how a
smaller value $\delta M\sim 100$ keV can improve the situation in
section IV.  In  section V we build a simple particle physics model
of DM  which can accommodate our findings, and also address the
PAMELA/ATIC/PPB-BETS observations.  Section VI gives a brief account
of the cosmological implications of the model.  Conclusions are given
in section VII.

\section{Methodology} To obtain the desired result for the 511 keV
signal, it is important to nearly saturate the
unitarity bound in at least one partial wave (or to have significant
scattering up to high values of $l$, but we will show that this does
not seem to be possible in the present context).  The Sommerfeld enhancement which can occur at low DM
velocities is crucial for getting such strong scattering.  We follow
the quantum mechanical treatment of appendix A.4 of ref.\ \cite{AH} 
to compute this effect.  

Because the gauge coupling $g\bar\chi_2 B_\mu
\gamma^\mu \chi_1$ is assumed to be off-diagonal, we have two states
$|1\rangle = |\chi_1,\chi_1\rangle$, 
$|2\rangle = |\chi_2,\chi_2\rangle$, whose interaction Hamiltonian has 
the matrix form
\beq
	V_{ij} = \left(\begin{array}{cc} 0 & -{\alpha e^{-\mu r}/r}\\
-{\alpha e^{-\mu r}/r} & 2\delta M \end{array}\right)
\label{Veq}
\eeq
where $\alpha = g^2/4\pi$ and $\mu$ is the mass of $B_\mu$. 
The wave function for the two-state system (with components labeled by
index $i$) in the CM frame
is $\Psi^i = \sum_{l}P_l(\cos\theta)R^i_{kl}(r)$, where $k$ is
the initial momentum.  Defining $\Phi_{l,i}(r) = R^i_{kl}/r$,
the Schr\"odinger equation is\footnote{For the numerical solution it is useful to rescale $r =
(\alpha/2\delta M) x$ and define the dimensionless variables 
$\Gamma = M_1\alpha^2/(2\delta M)$, $\Upsilon = (\alpha k/2\delta
M)^2$, $\eta = \alpha\mu/2\delta M$, so that the Schr\"odinger
equation takes the form $-\Phi'' +[l(l+1)/x^2 + \Gamma\hat V]\Phi = 
\Upsilon\Phi$, with the dimensionless potential $\hat V = ({0\atop
-e^{-\eta x}/x} {-e^{-\eta x}/x\atop 1})$.  We must have
$\Upsilon>\Gamma$ for the initial state to have enough energy to
produce the heavier $|\chi_2,\chi_2\rangle$ final state.
\label{fnote2}}
\beq
	-{1\over M_1}\Phi_{l,i}'' + \left({l(l+1)\over M_1 r^2}
	\delta_{ij}
	+ V_{ij}\right)\Phi_{l,j} = {k^2\over M_1} \Phi_{l,i}
\label{seq}
\eeq
The equation
is solved by the shooting method, where $\Phi_{l} \sim r^{l+1}
({1\atop b})$ near $r=0$ for some complex number $b$, which is then adjusted so
that there are only outgoing and not incoming waves in 
$\Phi_{l,2}$ as $r\to\infty$.

To extract the scattering amplitudes, we decompose the numerical
solution into incoming and outgoing waves, $\Phi_{l,1}^{\rm in}$,
$\Phi_{l,2}^{\rm in}$ and $\Phi_{l,2}^{\rm out.}$  Partial wave
unitarity implies the conservation of flux, $k|\Phi_{l,1}^{\rm
in}|^2 = k|\Phi_{l,1}^{\rm out}|^2 + k'|\Phi_{l,2}^{\rm out}|^2$
where $k'^2 \cong k^2 - 2 M_1\delta M$, which we use as a check on our
numerics.  The fraction of incoming $|\chi_1,\chi_1\rangle$ states
which gets converted to the $|\chi_2,\chi_2\rangle$ final state is
thus 
\beq
	f_l =  {k'\over k} {|\Phi_{l,2}^{\rm out}|^2\over 
	|\Phi_{l,1}^{\rm in}|^2}
\eeq
in the $l$th partial wave.  This must be integrated with the
Maxwell-Boltzmann distribution $Nv^2 e^{-v^2/v_0^2}$ (or some more
sophisticated distribution function, as we discuss below)
to find the thermally averaged cross section $\langle \sigma_{\rm gal}
v_{\rm rel}\rangle$.  Doing so modifies the unitarity-saturating
estimate (\ref{rl}) to read
\beq
	R_l \to 
4.5\times 10^{-4}\, {2l+1\over v_0}\left({500\ {\rm GeV}
	\over M_1}\right)^4\!\!\! \int_{u_t}^{u_{\rm esc}}\!\!\!\! 
	du\, e^{-u} f_l(u)
\label{rl2}
\eeq
where $u=v^2/v_0^2$, $v_t = \sqrt{2\delta M/M_1}$ is the threshold 
velocity for $\chi_2$
production and $v_{\rm esc}(r)\cong 700$ km/s is the escape velocity
at $r=0.4$ kpc, the outer edge of the region where INTEGRAL sees
excess $\gamma$ emission \cite{FW}.  Since $v_0 = 180$ km/s, the error in
extending the upper limit of integration to $\infty$ is small.  

\section{Constraints on large $\delta M$ model} We have manually scanned the parameter space of the model to try to
maximize the fraction of excited state particles, $f_l$, first
starting with the original class of models with the ``large'' mass
gap, $\delta M \cong m_e$.\footnote{In terms of ref.\ \cite{AH}, this 
actually requires looking at the scattering $\chi_0\chi_1 \to
\chi_1\chi_2$ since $\chi_1\chi_1 \to \chi_2\chi_2$ would have
$\delta M \approx 2 m_e$ and be even more suppressed.  Ref.\ \cite{PR}
has $\delta M\approx m_e$ by virtue of charged intermediate states,
which allows each incoming $\chi$ to be excited by only $m_e$ rather
than $2 m_e$.}\ \  
The goal here was to see if any parameters
could be found such that high partial waves could contribute, thus
overcoming the small prefactor $R_l \cong 3.4\times 10^{-3} (2l+1)$
at the optimal mass $M = m_e/2 v_0^2$ and $v_0=180$ km/s \cite{PR}.
We considered the limit in which the $B_\mu$ gauge boson mass
$\mu$ can be neglected.  Otherwise the range of the interaction
is reduced; this can only decrease the contribution from higher-$l$
partial waves, which correspond to scattering at large impact
parameter.  In that case, the relevant dimensionless 
parameter turns out to be
\beq
	\Gamma \equiv {M_1\alpha^2\over 2\delta M}
\label{geq}
\eeq
(see footnote \ref{fnote2}.)
For $\Gamma\gg 1$, the $s$-wave dominates the cross-section, while for
$\Gamma\ll 1$, a range of partial waves contribute significantly. 
However, in the latter case even the largest contributions fall 
short of the needed unitarity limit, so the total cross section is
not actually enhanced.  Working at the optimal mass $M=m_e/2 v_0^2$,
we have computed
\beq
I'\equiv e^{u_t}
\sum_{l}(2l+1)\int_{u_t}^{u_{\rm esc}}du\, e^{-u} f_l(u)
\label{Ieq}
\eeq
[see eq.\ (\ref{rl2})] as a
function of $\Gamma$.  This  quantity would take the
limiting value $\sum_l (2l+1)$ for all the partial waves which 
reach the unitarity limit $f_l=1$.  At the optimal mass 
$M= m_e/2 v_0^2$, $u_t = 4$,
and ref.\ \cite{PR} shows that $I'$ should have the value $(3.4\times
10^{-3})^{-1}\cong 300$ in order to match the INTEGRAL observations.
Fig.\ 3 shows that in the actual model, $I'$ reaches 
a maximum value of $0.3$ near $\Gamma=0.5$, far below what is needed.

\begin{figure}[t]
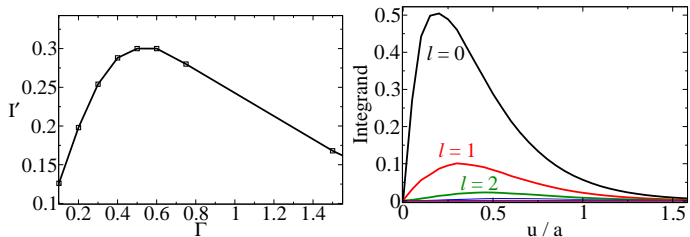
 %\smallskip 
\centerline{\epsfxsize=0.25\textwidth\epsfbox{IG-big.eps}
\epsfxsize=0.25\textwidth\epsfbox{integrand-big.eps}} \caption{Left:
Enhancement factor $I'$, eq.\ (\ref{Ieq}), as a function of $\Gamma
\equiv M_1\alpha^2/2\delta M$, in the disfavored case $\delta M =
m_e$, where the new gauge boson is taken to be massless and $M$ is
optimized.   The maximum enhancement is too small to match
observations. Right: integrands for successive partial waves
contributing  to $I'$, versus $u/a$, where $u = v^2/v_0^2$ and $a =
2\delta M/M v_0^2$.  One would need significant contributions from
many more partial waves to make $I'$ large enough.} \label{largedm}
\end{figure}

We have thus established that the high-$l$ loophole for the large $\delta M$
scenario does not work.  Another way of enhancing the effect
would be to take advantage of its strong dependence on $v_0$. One
would need to boost $v_0$ by the factor $(0.3\times 3.4\times
10^{-3})^{-1/7}  \cong 2.7$, giving $v_0 = 480$ km/s instead of 180
km/s.  On the other hand, more sophisticated estimates of the
dark matter distribution  indicate that $v_0$ is {\it
smaller} than the fiducial value, rather than larger, in the center
of the galaxy \cite{ldm2,PR}.  Different models of the DM distribution
function $n(r,v)$ give radially-dependent average velocities $v_0(r)$
which decrease toward $r=0$.  Therefore using the constant value for
$v_0$ which best describes the bulk of the galaxy already
overestimates the efficiency of DM excitation near the center, and 
the large $\delta M$ possibility seems to be ruled out (see however
footnote \ref{fnote1}). 

\section{The case of small $\delta M$}  Now we turn to the main
point, that smaller values of the DM mass splitting $\delta M\sim
100$ keV can overcome the problem of too small a signal, without any
need for increasing the DM velocity $v_0$.   Redoing the analysis of
ref.\ \cite{PR} for general $\delta M$, one finds that for the optimal
DM mass, $R_l$ is enhanced by an extra factor of $(m_e/\delta M)^4$.
We can therefore achieve the desired effect if unitarity is nearly
saturated only in the $s$-wave (or other low-$l$ contributions), with
$\delta M = 0.24\, m_e = 120$ keV.  
This  estimate applies for the optimal mass  $M = \delta M/2
v_0^2 = 170$ GeV.  However we can make the mechanism work at larger
$M$, as desired for getting the unified explanation of the
PAMELA/ATIC/PPB-BETS observations \cite{AH}, by making $\delta M$
only moderately smaller, as we will now show.

We have done a preliminary exploration of the parameter space, to see
what can be achieved in the concrete framework at hand. 
We defer a more comprehensive analysis to the future; here we will
just present a working example.  To get a large enough effect, it is
important to vary the mass $\mu$ of the exchanged gauge boson
$B_\mu$. Physically, this is due to resonant scattering when a bound
state of nearly zero energy forms \cite{AH,Hisano,MR}.  This effect is only
possible for  a finite-range potential such as the Yukawa type. 
Generally, we find enhanced scattering for larger values of $\Gamma=
{M_1\alpha^2/2\delta M}$, which is not surprising since the
interaction strength is $\alpha$, and there is an optimal (though not
sharply peaked) value of the dimensionless parameter 
\beq
	\eta \equiv
\alpha\mu/2\delta M
\eeq
giving the resonant effect. 

As an example, we present the case where $\Gamma = 10$, $\eta = 1.2$.
We consider the quantity $I\equiv e^{-u_t}I' = e^{-a} I'$ rather than
$I'$ of eq.\ (\ref{Ieq}), where $a \equiv 2 \delta M / (M v_0^2)$. In
the large $\delta M$ scenario, the best one could do was to optimize
$M$ such that $a=4$, but for small $\delta M$, we can obtain much
larger results at smaller values of $a$.  In contrast to the large
$\delta M$ case, we can hold $M$ fixed at a a value which is larger
than optimum explaining for the 511 keV line, but more interesting
for simultaneously explaining the other potentially DM-related
anomalies.  From eq.\ (\ref{rl2}) it follows that at $M=500$ GeV and
$v_0/c=0.0006$, one only needs  $I = 1.3$ to explain the INTEGRAL
observations.  

\begin{figure}[t]
\smallskip \centerline{\epsfxsize=0.45\textwidth\epsfbox{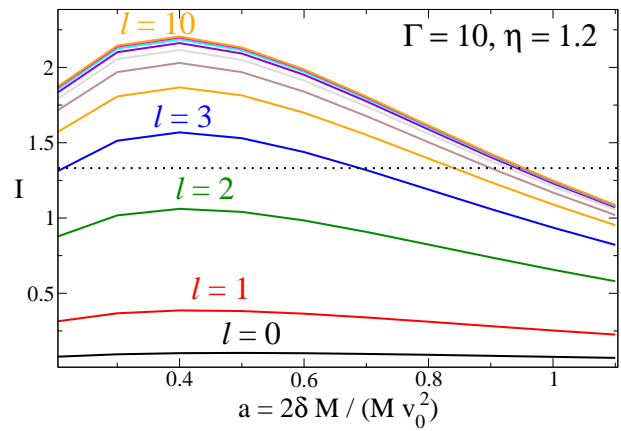}}
\caption{Enhancement factor $I\equiv e^{-u_t}I' = e^{-a} I'$ versus $a \equiv 2 \delta M / (M v_0^2)$
for the small $\delta M$ model, with $\Gamma=10$, $\eta=1.2$.  Successive contributions from partial
waves $l=0-10$ are shown.  $I=1.3$ (dotted line) is the value indicated by the
INTEGRAL observations.}
\label{adep}
\end{figure}

The enhancement factor $I$ is plotted as a function $a$ in figure
\ref{adep}, where the convergence of the successive partial wave
contributions is shown.  The needed value of $I=1.3$ can be obtained
for $a=0.95$.  For $M=500$ GeV and $v_0/c=0.0006$, this implies the
mass splitting $\delta M = 86$ keV.  From $\Gamma=10$ and $\eta=1.2$,
we infer that the fine-structure constant of the new  gauge
coupling is $\alpha = 0.0017$, and the mass of the gauge boson 
$B_\mu$ is $\mu = 120$ MeV.  

So far, we have assumed that the initial state $\chi_1$ has a
particular number density, namely that which was
 used in the analysis of
\cite{ldm2} to determine the cross section  $\sigma_{\rm gal} =
10^{-28}(M_1/{\rm TeV})^2$ cm$^2$ needed to explain the INTEGRAL
anomaly.  However, with three nearly degenerate DM states, one
expects the $\chi_1$ density to be $1/3$ this value, given that
$\chi_2$ decayed into $\chi_0$ and not $\chi_1$.  The signal is
proportional to the  integral of $\langle \sigma_{\rm gal} v_{\rm
rel}\rangle n^2(r)$ along our line of sight, so this would require
the cross section to be nine times higher than we have assumed. The
additional suppression can be counteracted if the DM density in the
central region $r<0.4$ kpc is three times higher than assumed in the
analysis of ref.\ \cite{ldm2}.  Even apart from any uncertainties in
the shape of $n(r)$, we note that it is normalized  to the local
energy density $n(8.5$ kpc) = $0.3$ GeV/cm$^3$, which is estimated to
be uncertain by a factor of 2 in the upward direction \cite{KK}.  If
this uncertainty works in our favor, we only need an additional
factor of $1.5$ enhancement in the central region.  This is a modest
shortfall, since in the four
models considered  by ref.\ \cite{ldm2}, $n(0.4$ kpc) varies by a
factor of 27.  Thus the smaller $\chi_1$ density does not seem to
pose a serious problem for the model.  On the other hand, the
missing factor of 300 in the large $\delta M$ case looks more
daunting.

We have also neglected the probable $r$-dependence of $v_0(r)$
discussed at the end of the previous section, which also tends to
reduce the predicted signal.  There is not yet a consensus on the
precise form of $v_0(r)$, but if it proves to give a significant
reduction, this can be compensated to some extent by taking more optimal
(smaller) values of the DM mass.  It is also possible that more
favorable examples at large $M_1$ exist, which would make it
interesting to perform a wider and more systematic search of the
model parameter space $(\Gamma,\ \eta)$ than we have been able to do
so far. 

\section{ATIC  and model building}

The simplest and most theoretically appealing way of getting three 
DM states $\chi_i$ and several gauge bosons is to assume that the
gauge symmetry is SU(2) and that $\chi_i$ transforms as a triplet. 
We will now explore the consequences of this hypothesis, with
particular attention to its implications for the  ATIC/PPB-BETS
excess $e^+e^-$ at 300-800 GeV.  Of course we will  also maintain all
ingredients needed for our successful explanation of the 511 keV
anomaly.

\begin{figure}[t] \smallskip
\centerline{\epsfxsize=0.45\textwidth\epsfbox{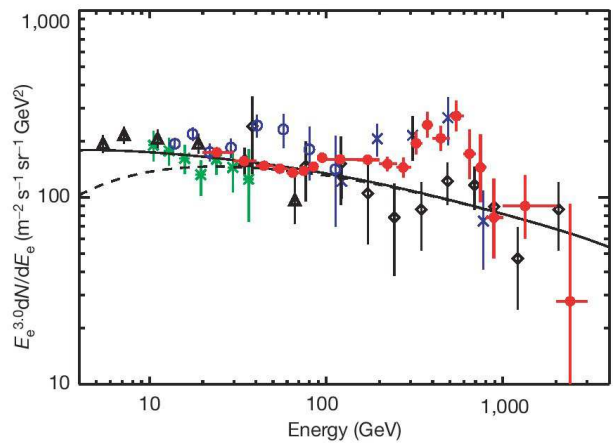}}
\caption{Energy spectrum of $e^+e^-$ observed by ATIC (solid circles),
taken from ref.\ \cite{atic}.
} \label{atic} \end{figure}

\begin{figure}[b] \smallskip
\centerline{\epsfxsize=0.25\textwidth\epsfbox{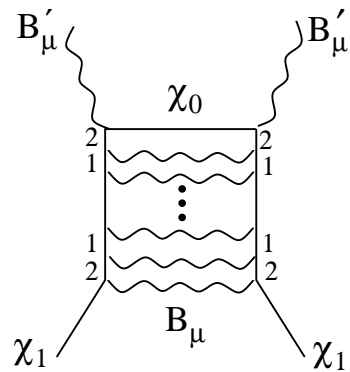}}
\caption{Sommerfeld-enhanced annihilation channel for explaining the
PAMELA/ATIC/PPB-BETS anomalies. Note that $B'$ will decay into $e^+e^-$.
The process suggested by
ref.\ \cite{AH} is shown. 
} \label{ann} \end{figure}

Already to explain INTEGRAL, we required two gauge bosons, $B$ which
couples to $\chi_1\chi_2$ to facilitate $\chi_1\chi_1\to\chi_2
\chi_2$, and $B'$ which mixes with the photon and mediates the decay
$\chi_2\to\chi_0 e^+ e^-$.   However, with only these states, it is
difficult to produce a rather sharply-peaked excess of high-energy
leptons on top of a lower energy continuum, which is suggested by the
ATIC data and which we have reproduced in figure \ref{atic}.   (The
PPB-BETS data are consistent, but not so strongly suggestive of the
peak.)  The process specified by ref.\ \cite{AH} in this regard is
shown in fig.\ \ref{ann}, where the final state $B'$ bosons
subsequently decay to $e^+ e^-$.  The  annihilation initially
produces  back-to-back $B's$, each carrying energy $M_1$, which
should give a leptonic spectrum  that is  rather uniformly
distributed in  energy.  It would be desirable to have an additional
channel which produces a single  nearly monoenergetic $e^+e^-$ pair,
each lepton having energy $M_1$.   Such a spike would be
significantly broadened by Coulomb scattering of the primary
particles in the galactic medium \cite{gil}-\cite{entropic}, possibly
giving the peak-like shape in the ATIC data.  We will come back to 
this issue in the next section.

Let us show how a simple SU(2) model could account for the
observations.  If $\chi_a$ is a real triplet, its gauge interactions are
\beq
	g\epsilon^{abc}\bar\chi_a B_b^\mu\gamma_\mu\chi_c
\label{epseq}
\eeq
Letting $a,b,c=0,1,2$, then in the previous notation, $B_0 = B$,
$B_1 = B'$, $B_2 = B''$.  A simple way to get kinetic mixing of the
photon with $B'$ is by including a Higgs triplet $\Delta_a$ in the
dark sector, and the dimension-5 operator
\beq
	{1\over\Lambda} Y^{\mu\nu} B^a_{\mu\nu}\Delta_a
\label{Bmixing}
\eeq
where $Y^{\mu\nu}$ is the field strength of the SM weak hypercharge
gauge field.  
If we assume that only $\Delta_1$ gets a VEV, this generates the 
required mixing of $B'$ with the photon and the $Z$ boson.  
It is
straightforward to work out the transformation that diagonalizes the
kinetic term of the $B'$, $A$ (photon) and $Z$ boson.  If $\mu'$ is the
mass of the $B'$, we find that the flavor states $B',A,Z$ are related
to the mass eigenstates $\tilde B',\tilde Y, \tilde Z$ by
\beqa
A &=& \tilde A - \epsilon \cos\theta_W \tilde B' +
O(\epsilon^2)\nonumber\\
        B' &=& \tilde B' + \epsilon \sin\theta_W \tilde Z 
+ O(\epsilon^2)\nonumber\\
        Z &=& \tilde Z - \epsilon \sin\theta_W {\mu'^2\over m^2_z}
	\tilde B'
 + O(\epsilon^2)
\eeqa
where $\epsilon\equiv 2\Delta_1/\Lambda$ and $\theta_W$ is the
Weinberg angle.  It is important for this class of
models that $B'$ does not mix with the photon $A$ at $O(\epsilon)$
[in fact neither does it mix at higher orders], because otherwise
$\chi_2\chi_0$ would acquire a coupling to $A$ which would be just
as strong as the coupling of $B'$ to $e^+e^-$.  In that case, INTEGRAL
would observe a narrow line from $\chi_2\to\chi_0\gamma$ at energy 
$\delta M\sim$ 1 MeV, in addition
to that at 511 keV, but this of course has not been seen.  The
$A$-$\tilde B'$ mixing implies a decay rate of $\tau^{-1}\sim 
\alpha_{\rm em}
\alpha M \epsilon^2 (m_e/\mu)^4$ for $\chi_2\to \chi_0 e^+ e^-$, and a
corresponding lifetime of order $10^{-5}$ s if $\epsilon\sim 10^{-4}$.
For  $\epsilon\sim 10^{-4}$ and $\langle\Delta_1\rangle
\sim10$ GeV as we will find below, the scale $\Lambda$ is $ \sim 50$ TeV.

With the breaking pattern $\langle\Delta_a\rangle =
\delta_{a1}\Delta$ needed in eq.\ (\ref{Bmixing}), 
the kinetic term $(D_\mu\Delta)^2$
of the triplet only gives mass terms for $B$ and $B''$,
\beq
	g^2 [\Delta_a\Delta_a B_b B_b - (\Delta_a B_a)^2] =
	g^2 \Delta^2 (B^2 + B''^2)
\label{dmass}
\eeq
(where $g$ is the dark SU(2) gauge coupling), so we must also include
symmetry breaking from another new Higgs field.  One simple
possibility is a symmetric traceless tensor, $\Sigma_{ab}$,
the 5D representation, whose VEV is only in the
$\Sigma_{02}=\Sigma_{20}$ or $\Sigma_{00}=-\Sigma_{22}$ components.
In fact, a global rotation around the $1$ direction (which is the 
subgroup of SU(2) left unbroken by our choice of $\langle\Delta_1\rangle$
as the triplet VEV) can conveniently put the 5-plet VEV in the
diagonal $\Sigma_{00}=-\Sigma_{22}$ components alone.  
The group generators in the 5D
representation can be written as $T^d_{ab,ce} =
i(\epsilon_{adc}\delta_{be} + \epsilon_{bde}\delta_{ac})$.  The VEV
$\Sigma_{00}=-\Sigma_{22}=\Sigma$
generates a mass term proportional to
\beq
	-g^2 B_d B_f \langle\Sigma_{ab}\rangle T^d_{ab,ce}T^f_{ce,hi}
	\langle\Sigma_{hi}\rangle = g^2\Sigma^2 (B^2 + B''^2 + 2B'^2)
\label{smass}
\eeq
The resulting gauge boson mass spectrum is then
$\mu = \mu'' = g\sqrt{\Sigma^2 + \Delta^2}$ and 
$\mu' = g\sqrt{2}\Sigma$.  
To get the desired
radiative mass splittings of the $\chi$'s below, we will need to
assume that $\Delta<\Sigma$, so that $\mu<\mu'$.

An important property of the interaction (\ref{epseq}) (and indeed all
the interactions in our model) is 
that it preserves the $Z_2$ symmetry $B\to-B$, $B''\to-B''$,
$\chi_1\to-\chi_1$ which is needed to keep $\chi_1$ stable, if
we assume that $\Delta_0$ and $\Delta_2$ are also charged under the
$Z_2$.  The discrete symmetry is thus unbroken by the VEV of
$\Delta_1$.  Note that  the  nonabelian cubic and quartic
interactions of the gauge bosons have the schematic form $B B'
B''$ and  $(B^2 + B''^2)B'^2$, which also respect the $Z_2$
symmetry.  A notable consequence of the symmetry is that $B,B''$ 
do not acquire any couplings to SM matter which would allow them to
decay into $e^+e^-$, nor to cause nuclear recoil in direct DM detection
experiments.  Therefore even though the excitation $\chi_1\to\chi_2$
has the right kinematics for the IDM explanation of DAMA, the
putatively exchanged $B$-$B''$ boson cannot interact with the
detector (nor can $\chi_0\to\chi_1$ work, both for this reason and
because of the larger mass splitting).   The remaining excitation 
$\chi_0\to\chi_2$ is possible from the point of view of the
interactions of the exchanged $B'$ boson, but the mass splitting
$\sim 2 m_e$ is too large for it to proceed at a detectable rate.

\begin{figure}[t] \smallskip
\centerline{\epsfxsize=0.45\textwidth\epsfbox{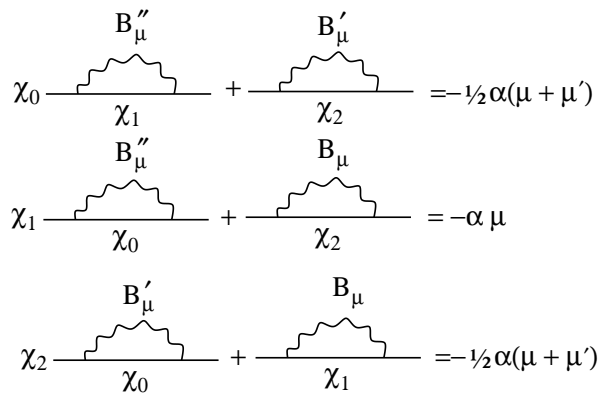}}
\caption{Radiative corrections to the $\chi_a$ masses, which leave
$\chi_0$ and $\chi_2$ degenerate.  $\mu$ and $\mu'$ are the masses of
the $B/B''$ and $B'$ gauge bosons, respectively.} 
\label{vacpol} \end{figure}

Next we consider the spectrum of the DM triplet.  
The tree level mass $M\bar\chi^a\chi_a$
gets split by $\frac12\alpha(\mu'-\mu)$ \cite{cheung} by the diagrams of fig.\
\ref{vacpol}.\footnote{It is important to put the external DM states on shell
to get the correct result.}\ \ 
We are assuming that $\mu<\mu'$ since, as we will show, 
this is what gives the desired
DM spectrum.   Figure  \ref{vacpol} indicates that
the radiative correction gives 
$\chi_1$ a mass which is larger than that of 
$\chi_0$ and $\chi_2$ by $\frac12\alpha(\mu'-\mu)$, but 
$\chi_0$ and $\chi_2$
but remain degenerate with each other.  We can break the remaining
degeneracy using the same VEV of the 5-plet as in eq.\
(\ref{smass}) by including the Yukawa
interaction
\beq
	h\Sigma_{ab}\bar\chi_a\chi_b
\label{triplet}
\eeq
This splits $\chi_{0,2}$ by $\pm h\Sigma$.
The resulting spectrum has the form
\beq
\left(\begin{array}{c}M_2\\ M_1\\ M_0\end{array}\right) = 
	M - \alpha\mu + 
\left(\begin{array}{c}h\Sigma -\frac12\alpha(\mu'-\mu)\\ 
	0\\ -h\Sigma -\frac12\alpha(\mu'-\mu)
\end{array}\right)
\eeq
We should choose $h\Sigma \gsim m_e$ to allow the decay of
$\chi_2\to\chi_0 e^+ e^-$, and $\frac12\alpha(\mu-\mu') \cong m_e
-\delta M$ to get the small splitting $\delta M$.  Using the value
$\alpha=1.7\times 10^{-3}$ suggested by our analysis of the INTEGRAL
signal, and assuming that $\Delta=\Sigma/2$ for example, we 
get the required spectrum with $\Sigma \cong 11.4$ GeV, 
$\Delta\cong 5.7$ GeV,  $h \cong 4.5\times 10^{-5}$.

We can easily construct a potential for the Higgs sector which
leads to  the desired symmetry breaking pattern:
\beqa
	V &=& \lambda_1(\sfrac12\Sigma_{ab}\Sigma_{ab}-\Sigma^2)^2
	+ \lambda_2(\Delta_a\Delta_a -\Delta^2)^2 \nonumber\\
	&+& \lambda_3 \Delta_a\Sigma_{ab}\Sigma_{bc}\Delta_c
\eeqa
To see that this can work, first consider the limit $\lambda_3=0$.
The VEV of $\Delta$ can always be rotated into the $\Delta_1$
direction by a global SU(2) transformation, while the VEV of
$\Sigma$ has no preferred orientation.  When $\lambda_3$ is turned on,
$\langle\Sigma_{ab}\rangle$ prefers to have the elements in the
first row and column vanish, 
$\langle\Sigma_{1a}\rangle= \langle\Sigma_{a1}\rangle=0$, so that the
$\lambda_3$ term remains zero.  We can still perform a global rotation
around the 1 axis to make $\langle\Sigma_{02}\rangle=
\langle\Sigma_{20}\rangle$ vanish; this rotation leaves
$\langle\Delta_a\rangle$ invariant.  

To summarize, this model is extremely simple: it needs a new SU(2)
(not SU(2)$\times$U(1) \cite{cheung}) in the dark sector,  broken by
a new Higgs triplet and 5-plet, and a dimension-5 coupling which induces mixing
between the photon and the new $B'$ gauge boson. These ingredients
easily give us the desired mass spectrum, fig.\ \ref{spect}(a), and
coupling of one of the gauge bosons $B'$ to leptons.  The stability
of the middle state $\chi_1$ is guaranteed by an unbroken $Z_2$
symmetry, which also keeps the 0.1 GeV-scale $B$ and $B''$ gauge
bosons stable, but these (as we will show presently) are
cosmologically harmless and not subject to accelerator constraints.

\section{Cosmological implications}

\begin{figure}[t] \smallskip
\centerline{\epsfxsize=0.45\textwidth\epsfbox{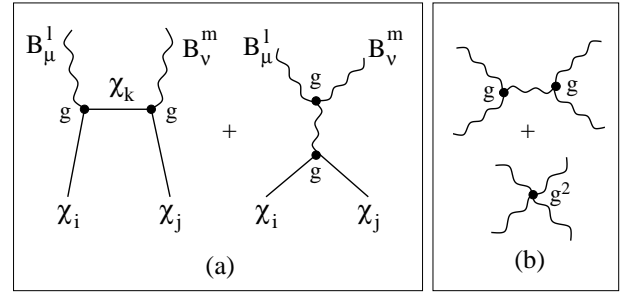}}
\caption{Left (a): diagrams determining the relic density of the DM
particles $\chi_i$.  Not shown is the $u$-channel version of the
first diagram.  Right (b): diagrams determining the relic density
of the new gauge bosons $B_i$.} 
\label{freezeout} \end{figure}

It is interesting to note that the model we have put forward based
primarily on the INTEGRAL and ATIC observations happens to predict a
relic density for the DM which is not so far from the required value.  To
match the WMAP value $\Omega_\chi h^2\cong 0.1$, one needs
$\langle\sigma_{\rm ann} v_{\rm rel}\rangle\cong 1$ pb$\cdot c$
\cite{PDG}; for three colors of nearly degenerate DM, this becomes
$3$ pb$\cdot c$.  Take $v_{\rm rel} = 2 p_{1}/M$, where $p_1$ is the
momentum of one of the incoming $\chi_i$'s in the center of mass
frame.  At the freeze-out temperature $T\sim M/20$, $p_{1}\sim
\sqrt{3TM}$.  The Mandelstam variable $t$ ranges between the values
$t_\pm \cong -M^2\pm 2Mp_1\equiv -M^2\pm\Delta t/2$, while $s\cong 4
M^2$.  A somewhat detailed computation gives us an estimate for
the matrix element for the annihilation process shown in 
fig.\ \ref{freezeout}(a) as $|{\cal M}|^2\cong 9 g^4$ (however we have
not been careful enough to determine whether the three diagrams
interfere constructively or destructively---this estimate assumes the
former).  Then 
\beq
	\sigma v_{\rm rel} \cong {|{\cal M}|^2 \Delta t\over 32 \pi s p_{1} M}
	 \cong {9\pi\alpha^2\over 2 M^2}
\label{sv}
\eeq

For the values we have favored thus far, $\alpha = 1.7\times 10^{-3}$
and $M=500$ Gev, (\ref{sv}) comes out too small by a factor of $\sim
100$.  We
should thus hope to find an example of sufficient positron production
for INTEGRAL at a value of $\alpha$ approximately 10 times higher,
$\alpha\cong 0.02$.  This is preferable to 
achieving the right relic density by making $M$ smaller, since
that would force
us to give up our explanation for ATIC.
As we mentioned before, the numerical scattering computation
becomes prohibitively slow at these larger values of
$\alpha$; work is in progress to explore this region of parameter
space.

\begin{figure}[b] \smallskip
\centerline{\epsfxsize=0.45\textwidth\epsfbox{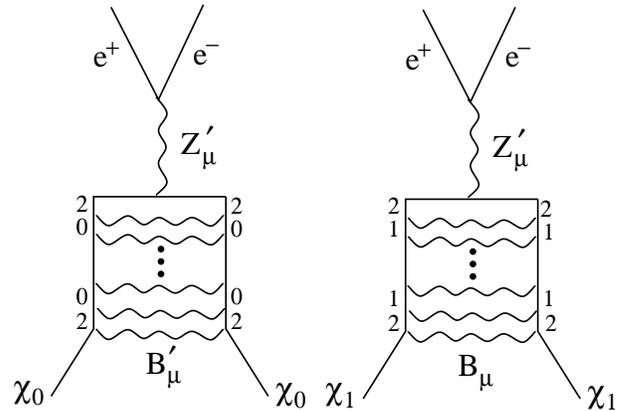}}
\caption{Processes which can give a sharper feature in the electron
spectrum, as suggested by the ATIC data.} 
\label{ann4} \end{figure}

However, it is possible to add extra interactions to the
model to adjust the relic density without the need for changing our
preferred value of the gauge coupling or DM mass.  If there is an
extra $Z'$  gauge boson with the couplings  $g_\chi \bar\chi_a
Z'_\mu\gamma^\mu\gamma_5\chi_a$ to the DM and  $g_e \bar e_R
Z'_\mu\gamma^\mu e_R$ to right-handed electrons, it can provide extra
annihilation channels which easily bring  $\langle\sigma_{\rm ann}
v_{\rm rel}\rangle$ up to the required value; we just need $g_\chi
g_e /M_{Z'}^2\cong 10 g^2/M^2$.  If the $Z'$ for some reason couples
mainly to $e_R$ and not other SM particles, this can maintain
the preference for annihilation into  $e^+e^-$ but not heavier
charged particles, as indicated by PAMELA/ATIC.  Such a $Z'$
would have to correspond to a U(1) symmetry broken at the scale
$M$ since the $\chi_i$ masses do not conserve the current to which
$Z'$ couples.  If such a $Z'$ solves the relic density problem, then its
contribution to annihilation in the galaxy will
also dominate the PAMELA/ATIC signals, giving nearly monoenergetic
$e^+e^-$ via the process of fig.\ \ref{ann4}.

The analogous freeze-out computation can be done for the diagrams of
fig.\ \ref{freezeout}(b) to determine the relic density of the $\mu
\cong
100$ MeV stable gauge bosons $B$-$B''$.  The estimate for
$\langle\sigma_{\rm ann} v_{\rm rel} \rangle$ is similar to to eq.\
(\ref{sv}), except that the relevant mass scale is $\mu$ rather than
$M$.  This makes the cross section larger by a factor of $10^7$. 
Since $\Omega$ scales like $1/\sigma$, the cosmological energy
density of the stable $B$'s is negligible, $\Omega_B \sim 10^{-6}$. 
They can annihilate into $e^+e^-$ because of their coupling to $B'$,
but this would probably be very difficult to detect due to the low
density of the $B$'s.  The stable gauge bosons thus seem to be rather
innocuous cosmological relics.

{\bf Note added.}  After completing the first version of this work,
M.\ Pospelov pointed out a potentially serious difficulty: the
process $\chi_1\chi_1\to\chi_0\chi_0$ can also proceed through
Sommerfeld-enhanced scatterings in the early universe, exponentially
depleting the density of $\chi_1$ states which we assumed to be  as
numerous as the $\chi_0$'s.  This would take place well below the
initial freeze-out temperature, when the kinetic energy of the
$\chi$'s becomes comparable to the mass splitting $\sim 2m_e$, and
their velocity is of order $\sqrt{2m_e/M_1}\sim 10^{-3}$, close to
the value in the galaxy today.  

To estimate the severity of the problem, we compute the rate of
$\chi_1\chi_1\to\chi_0\chi_0$  scatterings over the Hubble rate as a
function of temperature.  We take the value of the cross section
required for matching the INTEGRAL observation, $\langle\sigma v
\rangle \cong 10^{-28}(M_1/{\rm TeV})^2 (v_0/v)$ cm$^2 $(where
$v_0$ is the typical velocity in the galaxy), and the density
$n= \nu T^2$ with $\nu=10^{-12}\times(500{\rm\ GeV}/M_1)$
corresponding to the value  $\Omega_{DM} \cong 0.1$.  The DM
fell out of equilibrium at $T\cong M/20\cong 25$ GeV; at this time
its momentum was $p\sim\sqrt{3 M T} = M\sqrt{3/20}$, and thereafter
it redshifted like $T$.  We can write $p = c_p T$, where
$c_p \cong \sqrt{60}$ if the DM underwent standard freeze-out. 
However we will be interested in more general, nonthermal ways of
generating the DM, so we keep $c_p$ unspecified for now.   Taking
$H\cong \sqrt{g_*}T^2/M_p$, we find
\beq
	{n\langle\sigma v\rangle\over H} 
	= 4\times 10^{-5}{\nu M_1 M_p\over \sqrt{g_*} c_p
	{\rm GeV}^2} = {2\times 10^{5}\over \sqrt{g_*} c_p}
\label{ratio}
\eeq
This is independent of $T$ except through the number of species
$g_*$.  From (\ref{ratio}) it is clear that the $\chi_1$-depleting
process will be in equilibrium in the standard scenario, where
$c_p\cong\sqrt{60}$.  We need $c_p\gsim 10^5$ to avoid the depletion of
$\chi_1$.  This could happen if the $\chi$'s were produced
nonthermally.  A straightforward way would be through the late decays
of a heavy scalar $S\to\chi_i\chi_i$.  For example if $m_S\gsim 2 M_1$
and the decay occurs at $T\lsim 5$ MeV, then $p\sim M_1$ and $c_p\cong
M_1/T \sim 10^5$. This nonstandard production mechanism requires that
the standard one be subdominant, {\it i.e.,} we need larger values 
of $\alpha$ or of the $Z'$ couplings (as discussed above) to suppress
the relic density at freeze-out of the perturbative annihilations.  
It is straightforward to verify that the $\chi$'s will not reestablish
kinetic equilibrium with the SM particles after this point.  By
design, the strongest interaction with the SM is the scattering 
$\chi_i e^\pm \to \chi_i e^\pm$ via $t$-channel exchange of the $Z'$
boson.  The rate for this is 8 orders of magnitude smaller than the
Hubble rate at $T\sim 5$ MeV.

One question which arises with respect to out-of-equilibrium
production of DM through decays is how to get a sufficiently small
coupling between $S$ and $\chi$ so that the decay happens at such low
temperatures.  However, this can be naturally explained in the
context of a Grand Unified Theory (GUT), with a heavy gauge boson $X$
that couples to $S$ and to $\chi$.  Integrating out the $X$ boson
gives rise to an interaction of the form
$g_X^2|S|^2\bar\chi_a\chi^a/M_X$, where $g_X$ is the GUT gauge
coupling.  This interaction does not allow for $S$ to decay, but 
if we suppose that $S$ gets a VEV at the TeV scale, a coupling which
leads to decays is generated, $\lambda= g_X^2 \langle
S\rangle/M_X$.  The decay rate $\Gamma \cong 3\lambda^2 M_S /16\pi$
must equal the Hubble rate $H\cong g_* T^2/M_p$ at $T\lsim 5$ MeV, leading
to $\lambda\lsim  10^{-12}$ for $M_S\sim 1$ TeV.  Taking
$\langle S\rangle \sim 10$ TeV, we find $M_X/g^2_X \sim
10^{16}$ GeV, the GUT scale.  Of course $S$ would need to
have stronger couplings to other particles to freeze out at the right
density.

\section{Conclusions}

We have demonstrated a modification of the original XDM scenario
which makes the effect strong enough to explain the 511 keV anomaly,
without any radical change in our understanding of the DM
distribution in the galaxy, and within a theoretically sound class of
particle physics models.  The key idea is to have  two nearly
degenerate states of stable DM, such that the heavier one must first
be excited by the small energy $\delta M\sim 100$ keV into a third
state, in order to decay to the lowest state (fig.\  \ref{spect}). 
Moreover, we need  at least two kinds of new  0.1 GeV-scale gauge
bosons, $B_\mu$ and $B'_\mu$ (with a theoretical preference for  a
third,
$B''_\mu$),  such that $B_\mu$ and $B''_\mu$ mediate the excitation
$\chi_1\chi_1\to\chi_2\chi_2$, while $B'_\mu$ is responsible for the
decay $\chi_2\to \chi_0 e^+ e^-$.  A novelty of this scenario is that
the first excited DM state is stable, in addition to the ground
state.

Although our primary interest was to address the INTEGRAL
observation, we have also been motivated by the suggestion \cite{AH}
of a class of models  which can simultaneously explain the other
anomalies; thus we have focused on parameters where $M_1$ is at least
500 GeV.  Since the INTEGRAL signal scales like $M_1^{-4}$, we can
more easily accommodate it by relaxing this requirement.  To make
contact with ATIC/PPB-BETS, it may be desirable to find a model in
which $M_1$ is somewhat higher than the 500 GeV value we were able to
obtain thus far, since the ATIC data  indicate an excess going up to
800 GeV.   The fact that we did not yet find an example with $M_1 >
500$ GeV may just be due to the numerical challenges of our method,
where the integration time  to solve eq.\ (\ref{seq}) (subject to the
appropriate boundary conditions) becomes increasingly long as
$\Gamma$ (eq.\ (\ref{geq})) and $M_1$ are increased.   It would be
desirable to find an approximate analytic solution in the large
$\Gamma$ regime to make further progress.  

Our proposal for modifying the excited dark matter spectrum has also
motivated us to construct a model which is simpler  than others that
have been suggested along these lines \cite{cheung}; for example the
new gauge group in the dark sector is SU(2) rather than
SU(2)$\times$U(1).
A bonus of our model is the possibility of a new
signal due to annihilation of DM states into a single $e^+e^-$ pair
instead of two pairs, giving a sharper feature in the lepton spectrum
at energies near the DM mass.  In this paper we have not tried to
predict the PAMELA/ATIC/PPB-BETS signals in as much detail as we have
treated that of INTEGRAL; doing so would clearly be a worthwhile next
step.

$\phantom{AAA}$

We thank Nima Arkani-Hamed for interactions which revived our
interest in explaining the INTEGRAL anomaly, Gilles Gerbier for
enlightening insights about DAMA,  and Gil Holder for discussions on
DM density and electron energy loss,  Maxim Pospelov for disscussions
related to the ``note added,'' and Douglas Finkbeiner and Neal Weiner
for discussions about the LMXB hypothesis.  We also thank Itay Yavin for pointing out a sign error in the
radiative mass shifts in our first version.
\ \   FC is supported by a
Schulich Fellowship at McGill University. Our work is supported by
NSERC.

\end{document}